# A Machine-Independent Debugger—Revisited

David R. Hanson
Microsoft Research
`drh@microsoft.com`

January 1999

Technical Report
MSR-TR-99-4

## Abstract

Most debuggers are notoriously machine-dependent, but some recent research prototypes achieve varying degrees of machine-independence with novel designs. Cdb, a simple source-level debugger for C, is completely independent of its target architecture. This independence is achieved by embedding symbol tables and debugging code in the target program, which costs both time and space. This paper describes a revised design and implementation of cdb that reduces the space cost by nearly one-half and the time cost by 13% by storing symbol tables in external files. A symbol table is defined by a 31-line grammar in the Abstract Syntax Description Language (ASDL). ASDL is a domain-specific language for specifying tree data structures. The ASDL tools accept an ASDL grammar and generate code to construct, read, and write these data structures. Using ASDL automates implementing parts of the debugger, and the grammar documents the symbol table concisely. Using ASDL also suggested simplifications to the interface between the debugger and the target program. Perhaps most important, ASDL emphasizes that symbol tables are data structures, not file formats. Many of the pitfalls of working with low-level file formats can be avoided by focusing instead on high-level data structures and automating the implementation details.

# A Machine-Independent Debugger—Revisited

## Introduction

Historically, debuggers have been considered notoriously machine-dependent programs. Most debuggers in wide use today, e.g. the GNU debugger gdb [1] and Microsoft's Visual C++ debugger, do indeed depend heavily on a specific operating system or on a specific platform or compiler. There are, however, a few machine-independent debuggers; recent examples include ldb [2] and cdb [3], both of which are source-level debuggers for C.

Retargetable debuggers occupy a vast design space, because there are numerous axes of machine dependence. Debuggers can depend on machine architectures, operating systems, compilers, and linkers, and perhaps even user-interfaces. There are many ways to trade retargetability for functionality and efficiency, and different debuggers make different choices. For example, gdb provides a rich debugging repertoire and uses platform-specific symbol-table formats at the cost of requiring a substantial amount of platform-specific code. Ldb is easier to port to a different architecture, but it uses its own symbol-table format and thus requires cooperation from compilers.

Cdb explores perhaps the extreme reaches of this design space: It is nearly completely independent of architectures and operating systems, but it achieves this independence by loading a small amount of code with the target program and by having the compiler emit a non-standard (but machine-independent) symbol table. While these costs are non-trivial, cdb does illustrate how focusing on retargetability can simply a debugger dramatically.

This paper describes a new version of cdb that occupies another point in the debugger design space—with a twist. The major change in this version is that the symbol table is stored in a separate file instead of being embedded in the program executable file. Storing debugging information externally is increasingly common, because doing so makes it easier to store more information about the target program. For example, versions 5 and 6 of Microsoft's Visual C++ write debugging information to 'program database' files.

The twist is that the contents of the external symbol table are defined by a grammar written in the Abstract Syntax Description Language (ASDL) [4], and the ASDL tools generate the code that constructs, reads, and writes these symbol tables from this grammar. The resulting debugger is thus smaller and more reliable, because a significant amount of its code is generated automatically from a compact specification. ASDL is part of the National Compiler Infrastructure project, which seeks to reduce the costs of computer systems research by developing tools that make it easier and quicker to build high-quality, modular compilers. ASDL was designed for specifying the tree data structures often found in compiler intermediate representations, but it can specify tree data structures for any application, such as symbol tables for debuggers. Using ASDL with cdb suggests that ASDL is more widely applicable than originally envisioned.

## Background

A *nub* is the central feature of cdb's original and revised designs. As depicted in Figure 1, all communication between the target program and the debugger goes through the nub. The nub is a small amount of code loaded with the target; the debugger can be either in the same address space as the target, or in a separate address space. The latter configuration is the common one, because it protects the debugger from corruption by the target program.

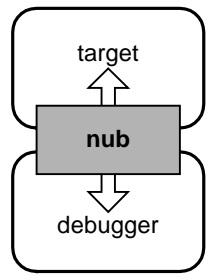

Figure 1. Nub-based debugger design.

## Interface

Interaction with the nub is defined by a small interface summarized in Figure 2. This interface is intentionally frugal because, while the interface itself is machine-independent, its implementations are not; an implementation for a spe-cific platform might depend on everything—small interfaces permit small implementations. For example, the nub used with cdb depends only the compiler (lcc [5]) and operating system (several Unix variants and Windows NT/95/98) and is less than 260 lines of C. The nub can be—and has been—used with other debuggers for other languages [6].

```
typedef struct {
        char file[32];
        unsigned short x, y
} Nub_coord_T;

typedef struct {
        char name[32];
        Nub_coord_T src;
        char *fp;
        void *context;
} Nub_state_T;

typedef void (*Nub_callback_T)(Nub_state_T state);

extern void _Nub_init(Nub_callback_T startup, Nub_callback_T fault);
extern void _Nub_src(Nub_coord_T src,
        void apply(int i, const Nub_coord_T *src, void *cl), void *cl);
extern Nub_callback_T _Nub_set(Nub_coord_T src, Nub_callback_T onbreak);
extern Nub_callback_T _Nub_remove(Nub_coord_T src);
extern int _Nub_fetch(int space, const void *address, void *buf, int nbytes);
extern int _Nub_store(int space, void *address, const void *buf, int nbytes);
extern int _Nub_frame(int n, Nub_state_T *state);
```

Figure 2. The nub interface.

The two data types and seven functions listed in Figure 2 permit a debugger to control a target and to read and write data from a target. The nub is mainly a conduit for opaque data; for example, it knows nothing about specific symbol-table formats, but provides simple mechanisms for reading them.

**_Nub_init** is called by the start-up code and initializes the nub; its arguments are pointers to callback functions that are called by the nub to initialize the debugger and to trap to the debugger when a fault occurs. As detailed below the type **Nub_state_T** describes the state of a stopped target, which occurs at start-up, breakpoints, and faults. **_Nub_set**, **_Nub_remove**, and **_Nub_src** collaborate to implement breakpoints. Stopping points define program locations at which breakpoints can be set in terms of 'source coordinates' specified by the type **Nub_coord_T**. A coordinate consists of a file name, a line number (**y**) and a character number in that line (**x**). The set of allowable stopping points depends on the language and the compiler; unlike most debuggers, which limit breakpoints to lines, cdb and lcc permit breakpoints to be set at any expression. **_Nub_src** enumerates the stopping points, calling the debugger-supplied **apply** function for each point, **_Nub_set** sets a breakpoint, and **_Nub_remove** removes one. When a breakpoint occurs, the breakpoint handler passed to **_Nub_set** as **onbreak** is called with a **Nub_state_T** value that describes the current state of the target.

**_Nub_fetch** and **_Nub_store** read and write bytes from the target's address spaces and return the number of bytes actually read and written. The target can have many abstract address spaces; one usually refers to the target's memory, while others refer to metadata about the target, including its symbol table. The compiler, debugger, and nub implementation define the conventions about address spaces; the nub interface specifies only a way to access those spaces.

Finally, **_Nub_frame** traverses the target's call stack; the top stack frame is numbered 0 and increasing numbers identify frames higher up the call chain. **_Nub_frame** moves to frame **n** and fills the **Nub_state_T** value with the state information describing that frame. The **fp** and **context** fields are opaque pointers that describe the target's state; these values are typically passed to **_Nub_fetch** to fetch symbol-table entries and the values of variables, for example.

## Implementation

The nub interface does not demand a machine-independent implementation. It is possible, for example, to provide an implementation that is specific to one architecture, operating system, and compilation environment. Cdb represents a different design choice: It sacrifices time and space for machine independence, and it runs on both UNIX and Windows platforms. Currently, it depends on the lcc compiler [5], because only that compiler emits the appropriate symbol tables and other debugging information, but those details are themselves compiler independent. Reference 3 gives the complete details of the implementation.

There are three major components to the cdb implementation: The nub, the compiler module that emits symbol tables and other debugging information, and the debugger itself, cdb. The debugger is a simple, command-line oriented debugger for C programs compiled by lcc. Programmers can set and remove breakpoints, examine the values of variables by name, and peruse the call stack. Cdb uses the nub interface to control the target program.

The compiler emits symbol tables as initialized C data structures, so their form is machine independent. The debugger reads these symbol tables by calling **_Nub_fetch** with an address space value that identifies the symbol table. Symbol-table entries include type and address information, so, given a symbol-table entry for a variable, cdb can fetch and display the values of the variable. Symbols are organized in an inverted tree according to scope; given a symbol, it and its ancestors are visible.

The compiler emits an array that identifies potential breakpoints, which the nub uses to implement **_Nub_set**, **_Nub_remove**, and **_Nub_src**. The compiler also emits code at each potential breakpoint that is essentially equivalent to the C expression

**(__module__V44b309d0.coordinates[$n$].i < 0 && _Nub_bp($n$, *tail*), *expr*)**

where **__module__V44b309d0** is a generated name and **coordinates** is the array of breakpoint data, $n$ is the stopping point number, *tail* is a pointer to the symbol-table entry for the last symbol in the scope in which the breakpoint appears, and *expr* is the C expression at the breakpoint. If the breakpoint is set, the **i** field of the array element is negative and **_Nub_bp** is invoked, which in turn calls the debugger's breakpoint callback function. Lcc injects this code in its intermediate representation, so the implementation is machine independent.

To implement **_Nub_frame**, lcc emits code to build a 'shadow stack' embedded in the normal call stack. This is accomplished by defining a local variable with a structure type for the shadow stack frame layout, and emitting code to link and unlink this frame at procedure entry and exit. There's no separate memory allocation involved, because the shadow stack frame is allocated along with other locals at procedure entry. The nub uses these frames to build **Nub_state_T** values, which are passed to the breakpoint and fault callback functions. Again, lcc emits these data and its associated code in its machine-independent intermediate representation.

Compared to other debuggers with similar capabilities, the cdb implementation is tiny: About 600 lines of C were added to lcc, and the nub, its associated communications functions, and cdb itself total about 1,500 lines (see Table 3, below). Of these, only the 31-line linking script, which sews together symbol tables from separately compiled modules, is potentially machine dependent; however, the same script has been used on all platforms to date.

## Revised Design

Cdb's machine-independence costs roughly a factor of 3–4 in both time and space. The symbol-table (including breakpoint data) and injected breakpoint code account for most of the space cost, and executing the injected breakpoint code accounts for most of the time cost. For example, when compiling itself with debugging on a SPARC, lcc emits about 749KB of symbol-table data and 1,347KB of injected code.

Much of the symbol-table data can be stored independently from the running code; the only important components are the addresses of the global variables and procedures. Thus, in the revised design, most of the symbol table is written to an external file. Newer compilers have adopted similar schemes; for example, Microsoft's Visual C++ most recent releases write symbol-table data to program database files. This approach can free the compiler from cramped, archaic symbol-table formats, and it makes symbol-table data available to other programs, such as source-code browsers.

Figure 3 illustrates this revised design. The nub interface remains unchanged, which is an important aspect of this design. Of course, the implementation is changed to accommodate external symbol-table files. The revised nub is divided into two components: The component that controls the target and accesses its address space is loaded with the target, as in the original design. The component that deals with the symbol table is loaded with the debugger, because, for the most part, this component deals only with the external symbol table. This revision is unnoticed by the debugger, because it sees the nub only through the unchanged nub interface. Figure 3 shows the two-process configuration, because it illustrates the two components of the nub most clearly. The situation is the same in the single-process configuration, but all four parts—target, debugger, and the two nub components—appear in the same address space.

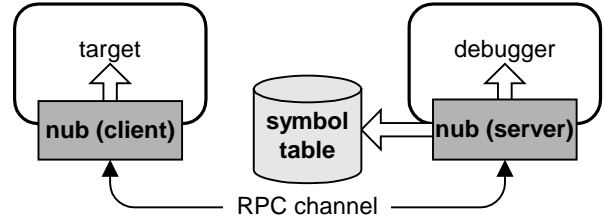

Figure 3. Revised design, in two-process mode.

Revising cdb to use an external symbol table requires a precise definition of the external representation, and functions to construct, read, and write the data. ASDL [4] automates much of this effort and asdlGen, the one of the ASDL tools, generates most of the necessary code.

## Implementation

As in the original implementation, lcc injects debugging code at the intermediate-representation level, but the injected code in the revised implementation is simpler, as detailed below. Instead of emitting symbol tables as initialized data structures embedded in the target program, lcc uses the ASDL-generated code to build the symbol table and write it to an external file.

The entire ASDL grammar is listed in Figure 4. The line numbers are for explanatory purposes only. ASDL is a domain-specific language for specifying tree data structures, and it's simple enough that it can be described easily by examples. An ASDL grammar is much like the definition of an algebraic data type. It consists of a sequence of ASDL productions that define an ASDL type by listing its constructor, the fields associated with each constructor, and the fields associated with all constructors for that type, which are called attributes. For example, lines 19–31 define 12 constructors for the ASDL type named '**type**', and the integer attributes **size** and **align**, which are common to all 12 constructors. The ASDL type **int** is a built-in type for integers.

This ASDL type represents C data types. The first four constructors (lines 19–22) define simple constructors for the basic C types; these constructors have no constructor-specific fields, only the common attributes. Lcc emits instances of these constructors for all of the C basic types. For example, on a 32-bit machine, the C type 'int' is represented with an instance of **INT** with a **size** and **align** both equal to 4, and an **INT** with a **size** and **align** equal to 1 represents the C type 'char'. The other basic C types are similarly represented.

Line 23 defines the constructor for C pointer types; it has one integer field (**type**) that identifies the referent type. ASDL grammars define trees, not graphs, so instances of ASDL types that are used more than once must be referenced indirectly. As described below, this ASDL grammar associates integers with instances of ASDL types that represent C types and C symbols. The constructors for arrays (line 27), functions (line 28), and qualified types (lines 29 and 30) also have integer fields that identify their referent types.

Line 24 defines **ENUM**, the constructor for C enumeration types. Its first field (**tag**) is an **identifier**, which is a built-in ASDL type, for the enumeration's C type tag. The second field (**ids**) is a sequence of **enum** types; the asterisk denotes a sequence. Line 18 defines **enum** as a record type with fields for the enumeration identifier and its associated value.

Structures and unions are defined similarly in lines 25 and 26. Both constructors carry the structure or union tag and a sequence of **field** records, which give the name, type, location of each C structure or union field. Bit fields are identified by nonzero values for **bitsize** and **lsb**. Function types (line 28) include a type for the return value (**type**) and a sequence of integers that identify the formal parameter types (**formals**).

AsdlGen generates all of the code necessary for constructing instances of the types defined in the grammar. To build an ASDL tree for a C type, lcc simply traverses its internal representation for the type and calls the appropriate generated functions. For example, given the C type declaration

```
enum color {RED=1,GREEN,BLUE};
```

lcc executes the equivalent of the following statement, assuming that enumeration types are implemented with 4-byte integers.

```
type = sym_ENUM(4, 4, "color",
        Seq_seq(sym_enum("RED", 1),
                sym_enum("GREEN", 2), sym_enum("BLUE", 3), NULL));
```

```
 1  module sym {
 2  module    = (identifier file,int uname,int nuids,
 3                item* items,int globals,spoint* spoints)
 4  spoint    = (coordinate src,int tail)
 5  item      = Symbol(symbol symbol)
 6            | Type(type type)
 7            attributes(int uid)
 8  coordinate   = (identifier file,int x,int y)
 9  symbol = STATIC(int index)
10          | GLOBAL(int index)
11          | TYPEDEF
12          | LOCAL(int offset)
13          | PARAM(int offset)
14          | ENUMCONST(int value)
15          attributes(identifier id,int uid,int module,
16                  coordinate src,int type,int uplink)
17  field     = (identifier id,int type,int offset,int bitsize,int lsb)
18  enum      = (identifier id,int value)
19  type      = INT
20            | UNSIGNED
21            | FLOAT
22            | VOID
23            | POINTER(int type)
24            | ENUM(identifier tag,enum* ids)
25            | STRUCT(identifier tag,field* fields)
26            | UNION(identifier tag,field* fields)
27            | ARRAY(int type,int nelems)
28            | FUNCTION(int type,int* formals)
29            | CONST(int type)
30            | VOLATILE(int type)
31            attributes(int size,int align)
32  }
```
Figure 4. ASDL grammar for the symbol table.

The actual code in lcc is nearly as simple as this example suggests; a single 75-line procedure handles all 12 constructors.

AsdlGen can generate code in C, C++, Java, ML, or Haskell, so clients can be written in whatever language best suits the application. Cdb is written in C.

## Symbol Tables

Lcc builds instances of the ASDL type **symbol** (Figure 4, lines 9–16) for each visible identifier. The constructors correspond to the different kinds of identifiers that appear in C programs. All symbols include the attributes defined in lines 15 and 16. The **id** field holds the symbol name itself, the **uid** field gives the symbol's unique identifying integer, uid for short, **type** holds the uid for the symbol's type, and **src** gives the location in the source program where the symbol is defined. As shown in line 8, a **coordinate** is a record that holds a file name (**file**), a line number (**y**), and a character number in that line (**x**).

The **uplink** field holds the uid for the previous symbol in the current scope or the last symbol in the enclosing scope. These fields form an inverted tree; given a **symbol** in this tree, that identifier and its ancestors comprise the set of visible identifiers in the compilation unit. Figure 5 shows the declaration fragments of a target program, **wf.c**, and the corresponding tree of **symbol**s. The arrows represent the **uplink** fields. For instance, if the target stops somewhere in the body of **getword**, the debugger will determine that the **symbol** for **c** identifies the set of visible symbols, which is given by following the arrows:

    **c s buf words main tprint getword isletter**

```
static int isletter(int c) { … }
static int getword(char *buf) { char *s; int c; … }
void tprint(struct node *tree) { … }
static struct node *words = NULL;
int main(int argc, char *argv[]) {char buf[40]; … }
```

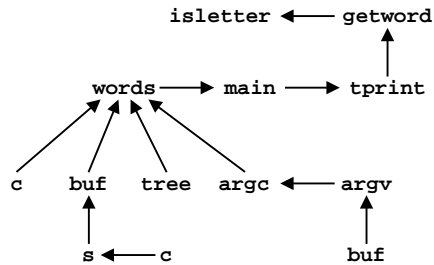

Figure 5. Sample program **wf.c** and its symbol table tree.

The **module** attribute is a unique integer name for the compilation unit in which the identifier appears. Global and static variables (**GLOBAL** and **STATIC**, lines 9 and 10) include the indices in the array of addresses described below, locals and parameters (**LOCAL** and **PARAM**, lines 12 and 13) include their offsets from the shadow stack frame, and enumeration constants (**ENUMCONST**, line 14) include the associate values.

Lcc wraps all of the symbol-table data into an instance of the ASDL type **module** defined in lines 2 and 3. This record starts with fields that give the file name of the compilation unit (**file**) and a unique integer name for the unit (**uname**) generated by lcc. This integer name is also used to generate the name of the external symbol-table file to which lcc writes the **module** using the procedure generated by asdlGen from the ASDL grammar.

A module also includes a sequence of **item** instances, which associate a **symbol** or **type** with a uid (lines 5–7), and the uid of the last global or static variable (the **global** field). For example, the **global** field in the **module** for the code in Figure 5 would contain the uid for **words**. The global fields are used for traversing all globals and statics in all compilation units during symbol-table searches.

The external symbol table contains everything about program identifiers except the addresses of globals (including functions), which are unknown until link time. Lcc emits into the target program an instance of the C type

```
struct module {
    unsigned int uname;
    void **addresses;
};
```

where **uname** is initialized to the integer name for the compilation unit and **addresses** is initialized to an array of addresses of the global identifiers defined in the unit. For example, lcc emits into read-only memory the equivalent of the following C fragments for Figure 5's **wf.c**.

```
const struct module _module_V49499895 = {
    0x49499895;
    &L93;
};
const void *L93[] = { &words, main, tprint, getword, isletter };
```

The variable **_module_V494999f8** includes the module's unique integer name. At link-time, a script scans all object files for names of this form and generates an initialized array of pointers to the module structures; for example, if **wf.c** is compiled with **lookup.c** and the resulting object files are linked together, the linking script generates the following code.

```
extern struct module _module_V49499895, _module_V494999f8;
const struct module *_Nub_modules[] = {
    &_module_V49499895;
    &_module_V494999f8;
    0
};
char _Nub_bpflags[37];
```

The object file for this code is compiled and loaded with the target program along with the nub. **_Nub_modules** gives the nub access to all of the **module** structures and thus to the files holding the symbol tables for all of the separately compiled C source files. The debugger uses **_Nub_fetch** to read the **module** structures and the address arrays. **_Nub_bpflags** is described in the next section.

## Breakpoints

The last field in a **module**, **spoints** (Figure 4, line 3), is a sequence of **spoint** records (line 4) that maps stopping points, which are the sequence element indices, to source coordinates and gives the uid for the symbol-table 'tail' for each stopping point. These data are used to implement **_Nub_set**, **_Nub_remove**, and **_Nub_src** and to supply the uid for the appropriate **symbol** when a breakpoint occurs. Cdb can set breakpoints at any individual expression and on the entry and exit points of compound statements. For example, the italicized superscripts in Figure 6 identify the stopping points in **getword**. Notice that it's possible to set a breakpoint on the right operand of the short-circuit AND operator, **&&**.

```
static int getword(char *buf) {⁸
    char *s;
    int c;
    while (⁹(c = getchar()) != -1 && ¹⁰isletter(c) == 0)
        ¹¹;
    for (¹²s = buf; ¹³(c = isletter(c)) != 0; ¹⁴c = getchar())
        ¹⁵*s++ = c;
    ¹⁶*s = 0;
    if (¹⁷s > buf)
        return ¹⁸1;
    return ¹⁹0;
}
```

Figure 6. Stopping points.

The compiler emits code at each stopping point that is essentially equivalent to the C expression

   **(_Nub_bpflags[$n$] != 0 && _Nub_bp($n$),** *expr***)**

where *expr* is the C expression at the stopping point $n$. For example, lcc emits

   **if ( (_Nub_bpflags[17] != 0 && _Nub_bp(17), s > buf) )**

for the if statement containing stopping point 17 in Figure 6. **_Nub_set** plants a breakpoint at a given source coordinate by searching the **spoints** sequence for the coordinate; if it's found, **_Nub_set** writes a one to the corresponding index in **_Nub_bpflags**. **_Nub_remove** clears an element in **_Nub_bpflags**. **_Nub_set** and **_Nub_remove** are implemented in the server (debugger) side of the nub, and they use **_Nub_store** with a distinguished address space identifier to write **_Nub_bpflags**, which is in the client (target) side of the nub.

When a breakpoint occurs, **_Nub_bp** uses the stopping point number to initialize the fields of the **Nub_state_T** value it passes to the debugger's call back function. The stopping point number leads to the source coordinate for the stopping point, the name of the function in which that point appears, and the uid of **symbol** that represents the set of visible identifiers at that point. This uid is used to set the **context** field in the **Nub_state_T** value to the appropriate **symbol**. The **fp** field is set to the appropriate shadow stack frame, as described below.

A given stopping point, say 17, can appear in every separately compiled module. Thus, **_Nub_bp** can be called at a non-existent breakpoint. The client-side nub passes this event on to the server-side nub, which simply dismisses extraneous breakpoints. This scheme simplifies the client-side nub at the cost of recognizing these occasional extraneous events. It also permits the module with the most stopping points to determine the size of **_Nub_bpflags**, which is generated at link-time. In the original cdb implementation, each separately compiled module included an array whose length was the number of stopping points in that module.

## Stack Frames

The nub must understand stack frames just enough to implement **_Nub_frame** and to provide an appropriate context for addressing parameters and locals. It does not, however, have to provide access to the machine-dependent details of the stack, because the interface provides no way to access or to use them. As in the original implementation, the revised implementation uses a shadow stack embedded in the normal call stack, but the revised frames are simpler and smaller. At function entry, the compiler generates a local variable by simulating the following declaration where **tos** is a generated name.

```
struct sframe {
    struct sframe *up, *down;
    int func;
    int module;
    int ip;
} tos;
```

There's no separate allocation required for **tos**; it's allocated along with other locals during function entry. Lcc also emits code to initialize the fields; e.g., for **getword**, it emits:

```
tos.down = _Nub_tos;
tos.func = 2;
tos.module = 0x49499895;
_Nub_tos = &tos;
```

The nub's private global **_Nub_tos** always points to the top frame on the shadow stack. The **down** field points to the previous shadow stack frame, the **func** field is the uid of the function, and the **module** field is the unique integer name for the compilation unit. The **ip** field is set the stopping point number when a breakpoint occurs and just before calls. The **up** field points up the shadow stack and is used only by **_Nub_frame**, which initializes this field only when necessary. Lcc also emits code at calls to set the **ip** field and at returns to pop the shadow stack by assigning **tos.down** to **_Nub_tos**.

Given a shadow stack frame, **_Nub_bp** builds a **Nub_state_T** value, as described above. The offsets stored in **PARAM** and **LOCAL symbol** values (see Figure 4, lines 12 and 13) are offsets from the shadow stack frame. All of these computations are done at the intermediate-code level, and are they independent of the target machine. They do, however, depend on some of the details of lcc's code generator architecture; for example, the offsets are computed by accessing code-generator data structures.

## Overhead

The original version of cdb achieved its machine independence at a cost of a factor of 3–4 in both space and time. The revised version reduces the space cost by nearly one-half and the time cost by about 13% without compromising machine independence. Table 1 lists the sizes in kilobytes of the code, initialized data, and uninitialized data segments for four variants of lcc version 4.1 and the sizes of the executable files themselves. The first three rows reproduce the measurements from the original version of cdb described in Reference 3. (These numbers are different than those shown in Reference 3 because the platform used in gathering the original measurements is no longer available and the original table used lcc version 3.5.) The fourth row lists the measurements for the revised version. All of the measurements in this section were taken on an SPARC Ultra 1/170 running SunOS 5.5.1 (Solaris 2.5.1) with 128MB of memory.

Table 1. Sizes in kilobytes of lcc executables.

| Code segment, KB | Initialized data, KB | Uninitialized data, KB | File size, KB | lcc variant |
|---|---|---|---|---|
| 504 | 28 | 37 | 559 | No debugging data |
| 504 | 28 | 37 | 1,249 | SunOS-specific debugger data |
| 2,065 | 563 | 37 | 2,662 | Original cdb debugging data and code |
| 1,651 | 29 | 39 | 1,717 | Revised cdb debugging data and code |
|  |  |  | 1,002 | External symbol-table files |
|  |  |  | 281 | Compressed external symbol table files |

In each row, the sum of the first three columns reveals the space overhead when subtracted from the sum for the first row; for example, the space overhead in the original version is 2,065+563+37–(504+28+37) = 2,096KB. Table 2 lists the sizes of the symbol-table portion of this overhead for both the original and revised versions. Of the 2,096KB in the original version, symbol-table data account for 749KB, so the injected debugging code accounts for the other 1,347KB.

For the revised version, the space overhead is 1,150KB, of which about 8KB is debugging data and 1,142KB is injected code. Of course, most of the symbol-table data in the revised version is written to external files; as shown in the last two lines Table 1 these files hold 1,002KB of data uncompressed. These data compress well using general-purpose compressors: zip, for example, compresses the 1,002KB of data into a 281KB zip file.

Table 2. Sizes of debugging data structures.

| Original, bytes | Revised, bytes | Data structure |
|---|---|---|
| 220,078 | | File names and identifiers |
| 226,772 | | Symbols |
| 209,996 | | Types |
| 109,236 | | Source coordinates |
| 396 | 264 | Modules |
| 392 | | Pointers to file names |
| | 4,008 | Address vectors |
| | 3,966 | Breakpoint flags |
| 766,870 | 8,238 | Total |

The display below gives the execution time in seconds for each of the variants of lcc listed in Table 1 to compile lcc 4.1, which amounts to 55,000+ lines of C after preprocessing. These times do not include preprocessing, assembly or linking.

| | |
|---|---|
| 14.5 secs. | No debugging data |
| 13.7 | Solaris-specific debugging data |
| 37.3 | Original cdb debugging data and code |
| 32.2 | Revised cdb debugging data and code |

As the last two lines indicate, lcc emits the revised symbol-table data a bit faster than it emits the original symbol-tables. When lcc emits the revised data, it's writing almost as much data as in the original implementation, but the data is written to external files instead of to the generated assembly-language files. It's a bit faster to write the external files because the writers generated by asdlGen are simpler than the assembly-language emitters in lcc.

## Discussion

Table 3 lists the source files for both the original and revised versions of cdb, gives their sizes in non-blank lines of source code, and summarizes their purposes. Almost every file in the revised version is shorter, because the ASDL-defined symbol tables are simpler. It's easier to emit these symbol tables, so the revised **stab.c**, the symbol-table emitter in lcc, is shorter; it's easier to process these tables in response to debugger commands, so **cdb.c**, the debugger proper, is shorter. Symbol-table interrogation (in **symtab.c**) requires a few more lines because, given a **symbol**, a search is required to find the **module** in which it is defined. The nub is longer because splitting it into two parts (**nub.c** and **nub2.c**) required some additional code to coordinate the interaction between them.

These line counts do not include the code generated by asdlGen from the ASDL grammar. The 31-line ASDL grammar generates about 1,320 lines of C declarations and function definitions. This code is approximately what would be required if the ASDL-generated constructors, readers, writers were written by hand. The external file format used by ASDL is independent of both language and platform, so the symbol tables can be read and written by other tools written in different ASDL-supported languages. The savings that were realized by using ASDL would thus increase if the ASDL grammar were used to generate code in other languages. For example, if a source-code browser were written in Java, it would use the 1,557 lines of Java generated from **sym.asdl**.

Table 3. Source code for original and revised implementations.

| Original | Revised | Module | Purpose |
|---|---|---|---|
| 31 | 39 | **prelink.sh** | linking script, per platform |
| 565 | 412 | **stab.c** | symbol-table and breakpoint code emitter |
| 249 | 125 | **nub.c** | the nub |
|  | 135 | **nub2.c** | debugger side of the revised nub |
| 191 | 164 | **client.c** | target side of the two-process nub |
| 202 | 155 | **server.c** | debugger side of the two-process nub |
|  | 52 | **comm.c** | common communication code |
| 794 | 682 | **cdb.c** | cdb's user interface and command processor |
| 80 | 124 | **symtab.c** | symbol-table and type management |
| 15 |  | **symstub.c** | `symtab` stubs for single-process debugger |
|  | 31 | **sym.asdl** | ASDL grammar |
| 2,127 | 1,919 |  | Total |

The nub interface was designed to be as simple as possible, but the experience with ASDL suggests a further simplification: The nub interface does not need high-level source coordinates. Debuggers must have a way to specify the locations for breakpoints, but a low-level mechanism is sufficient. In the revised design, the debugger side of the nub uses the symbol tables to map source coordinates to stopping-point numbers. The debugger could do this mapping directly, and the nub interface could be simplified to use only stopping-point numbers. Indeed, these numbers can be whatever is most convenient for the compiler and the nub implementation and thus could be treated as opaque integers in the interface. In implementations like cdb, they would be indices; in other implementations, they could be addresses or byte offsets from procedure entry points. Simplifying the nub interface would make room for other, more important, additions, such as support for single stepping [3].

Most systems view symbol tables as file formats and document them as such in torturously detailed specifications. (Similar comments apply to object code.) These arcane and archaic formats are nearly impervious to change. For example, most debuggers can set breakpoints only on lines, because the symbol-table format provides information only about lines even though the syntax of most languages is not line-oriented and includes operations that have embedded flow of control. This trend continues: Java's class files are described as a file format, and class files include metadata that maps locations to line numbers [7]. Using ASDL emphasizes that symbol tables are data structures, not file formats. ASDL grammars provide concise, abstract documentation for these data structures that focuses on their content, not their persistent format, and generating the code for readers and writers automates an error-prone aspect of system implementation and copes with change more gracefully.